# Simulation of ASTROD I test mass charging due to solar energetic particles

L. Liu[a,b], G. Bao[a,b], W.-T. Ni[a] and D.N.A. Shaul[c]

[a] Purple Mountain Observatory, Chinese Academy of Sciences, Nanjing,, 210008 China
[b] Graduate University of the Chinese Academy of Sciences, Beijing, 100049 China,
[c] Blackett Laboratory, Imperial College, London, SW7 2AZ, UK
Email address: lliu@pmo.ac.cn (L. Liu)

## Abstract

As ASTROD I travels through space, its test mass will accrue charge due to galactic cosmic-rays and solar energetic particles incident on the spacecraft. This test mass charge will result in Coulomb forces between the test mass and the surrounding electrodes. In earlier work using the GEANT4 toolkit, we predicted a net charging rate of nearly 9.0 +e/s from cosmic-ray protons between 0.1 and 1000 GeV at solar maximum, and rising to 26.5 +e/s at solar minimum. Here we use GEANT4 to simulate the charging process due to solar energetic particle events and to estimate the magnitude of acceleration noise due to this charging. The predicted charging rates range from 2840 to 64300 +e/s, at peak intensity, for the 4 largest SEP events in September and October 1989. For the 2 larger events, the acceleration disturbances due to charging exceeds the ASTROD I acceleration noise budget requirement. Continuous discharge should be considered for suppressing this charging noise. The acceleration noise during the 2 small events is well below the design target, although during these events, the net charging rate will be dominated by these solar fluxes.

## 1. Introduction

The ASTROD I mission concept is based around a single, drag-free spacecraft and laser interferometric ranging and pulse ranging with ground stations. It is the first step towards realising the ASTROD mission (the Astrodynamical Space Test of Relativity using Optical Devices) (Ni, 2006; Ni et al., 2004, 2006a, 2006b). The scientific goals of ASTROD I include measuring relativistic parameters with better accuracy, improving the sensitivity achieved in using the optical Doppler tracking method for detecting gravitational waves, and measuring many solar system parameters more precisely. Key to realizing these goals is ensuring that all forces, other than gravity, acting on the test mass (TM), are reduced to negligible levels.

Galactic cosmic rays (GCRs) and solar energetic particles (SEPs) can penetrate the shielding provided by the spacecraft (SC) and deposit electrical charge on the test mass. This charge will interact with the surrounding electrodes, resulting in spurious Coulomb forces on the TM. Our previous work predicted the charging rates for ASTROD I test mass from GCRs at solar minimum and at solar maximum using first simplified geometry (Bao et al., 2006a) and subsequently using a more realistic geometry model (Bao et al., 2006b, 2006c), and estimated the magnitude of disturbances associated with charging (Bao et al., 2006a, 2006b, 2006c). In this paper, we present a detailed calculation of ASTROD I test mass charging due to protons generated by SEP events, and compare this with charging due to primary protons at solar minimum and solar maximum (Bao et al., 2006b, 2006c). Using these results, we estimate and discuss the magnitude of the associated disturbances.

## 2. Modeling the charging process

*2.1 Geometry model*

We use ASTROD I geometry model from our previous GEANT4 work (Bao et al., 2006b, 2006c). Fig. 1 shows a sketch of this model. The basic spacecraft structure is a cylinder of 2.5 m diameter, 2 m height and 10 mm thickness. All surfaces of spacecraft are covered with a thermal shield, consisting of five layers of materials, which serves as a sunlight shield and reduces temperature perturbations. The top and bottom of the spacecraft are covered by the upper deck and lower deck and the side surface by solar panels. In orbit, the cylindrical axis of spacecraft will be perpendicular to the orbital plane, with the telescope pointing toward the ground laser station. The edges of upper deck and thermal shield are shown as large ellipses in Fig. 1. The inner lower deck is shown as the grey part in Fig. 1. The payload structure is used for shielding the optical bench, the inertial sensor, the telescope, etc. The telescope, which collects the incoming light, is a 500 mm diameter f/1 Cassegrain telescope. Some 30 boxes represent the components above ~ 0.1 kg mass, based on the LISA GEANT4 model figures (Araújo et al., 2003, 2005). The masses of the spacecraft and payload are estimated to be 341 kg and 109 kg in this study.

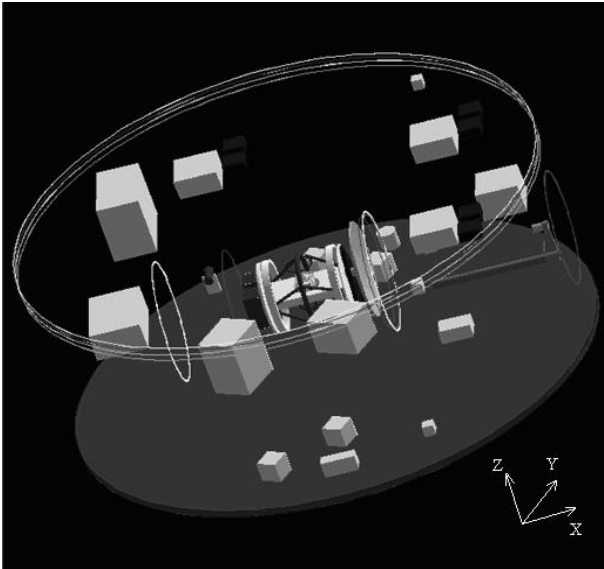

Fig. 1. The schematic diagram for the geometry models of ASTROD I spacecraft.

A $50 \times 50 \times 35$ mm$^3$ rectangular parallelepiped test mass made from Au-Pt alloy of low magnetic susceptibility ($< 5\times10^{-5}$) is at the centre of the spacecraft, surrounded on its all six faces by electrodes. The test mass is inside a molybdenum housing located in optical bench mounted behind the telescope. The test mass is surrounded by sensing and actuation electrodes fixed in molybdenum housing. A 0.3 μm gold layer is plated on the entire inner surface of the sensor housing. The assembly is placed inside a titanium vacuum ($< 10$ μPa) enclosure. The gap between test mass and electrodes along X axis and Y axis is 4 mm, and along Z axis 2 mm (Bao et al., 2006b). The GEANT4 model for the inertial sensor (IS) is shown in Fig. 2.

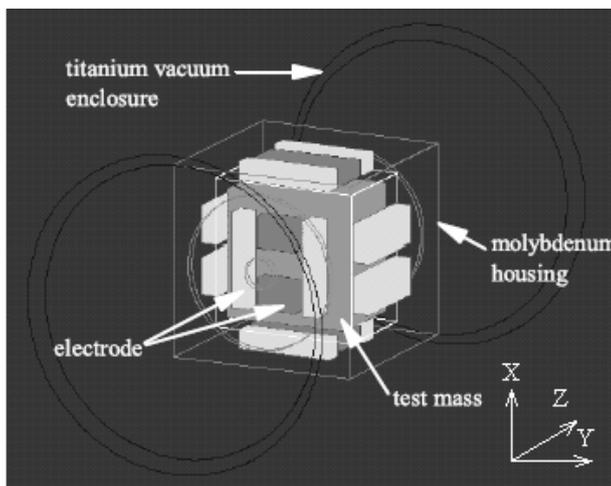

Fig. 2. ASTROD I inertial sensor model implemented in Geant4. The test mass, located at the centre of the figure, is surrounded by sensing electrodes (white) and injection electrodes (grey).

## 2.2 Solar energetic particles (SEPs)

SEPs have energies larger than 1 MeV and there are essentially two event classes:

impulsive and gradual. Particles generated by impulsive events show maximum energies of 50 MeV. Gradual events are associated with shock acceleration in coronal mass ejections (CMEs) and their spectrum can extend up to a few GeV energies (Vocca et al., 2005). In gradual SEP events, protons of energy greater than 100 MeV can penetrate spacecraft walls and payload structure, and penetrate the test mass. The events can have rapid onsets and durations from several hours to several days. Although much rarer than solar flares, SEP events can have large particle fluences and deposit significant amounts of charge over periods of days. It is therefore important to calculate their effects on ASTROD I test mass charging.

We took four example events that occurred on September 29, October 19, October 22 and October 24, 1989. The respective proton fluxes in three channels recorded by the Geostationary Operational Environmental Satellites (GOES) (http://www.ngdc.noaa.gov/) are shown in Fig. 3, Fig. 4, Fig. 5 and Fig. 6.

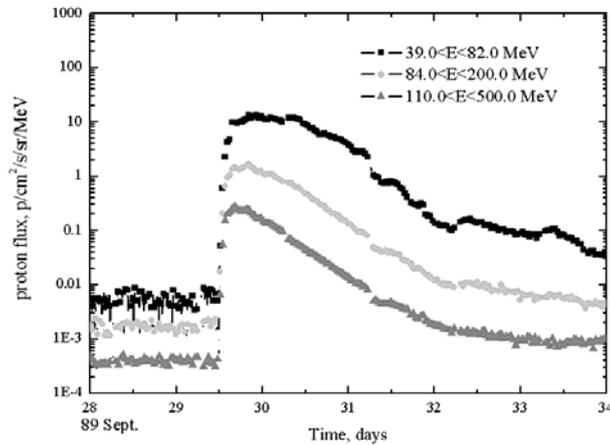

Fig. 3. Proton fluxes as a function of epoch in three energy ranges recorded by GOES for SEP event on 29 Sept. 1989.

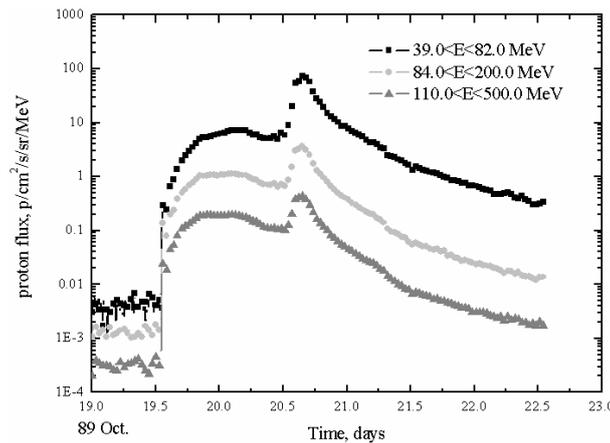

Fig. 4. Proton fluxes as a function of epoch in three energy ranges recorded by GOES for SEP event on 19 Oct. 1989.

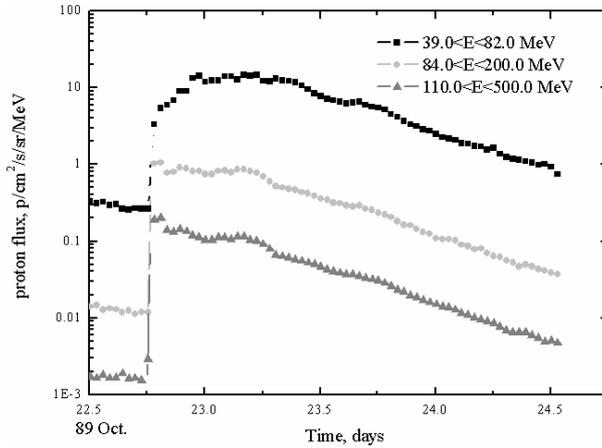
Fig. 5. Proton fluxes as a function of epoch in three energy ranges recorded by GOES for SEP event on 22 Oct. 1989.

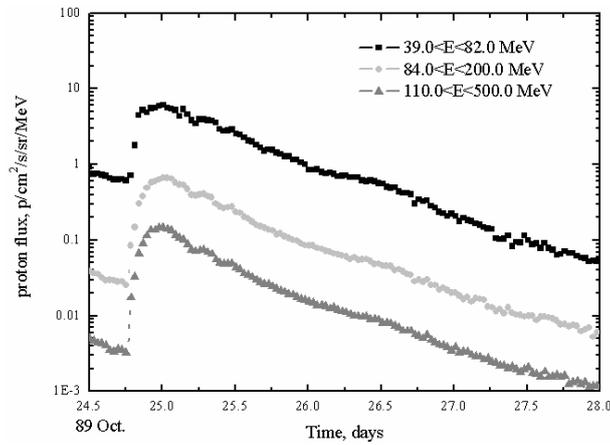
Fig. 6. Proton fluxes as a function of epoch in three energy ranges recorded by GOES for SEP event on 24 Oct. 1989.

The Weibull function is found to describe many SEP energy spectra (Xapsos et al., 2000). The Weibull differential flux (in protons/cm$^2$/s/sr/MeV) is given by,

$$\phi(E) = Ak\alpha E^{\alpha-1} \exp(-kE^\alpha), \tag{1}$$

where $E$ is the energy and $A$, $k$ and $\alpha$ are parameters. The Weibull parameters for the peak differential flux spectra are given by C.S. Dyer et al. (2003). Differential energy spectra at the peaks of the events are compared with the cosmic ray spectrum at solar maximum and solar minimum in Fig. 7 (Grimani et al., 2004).

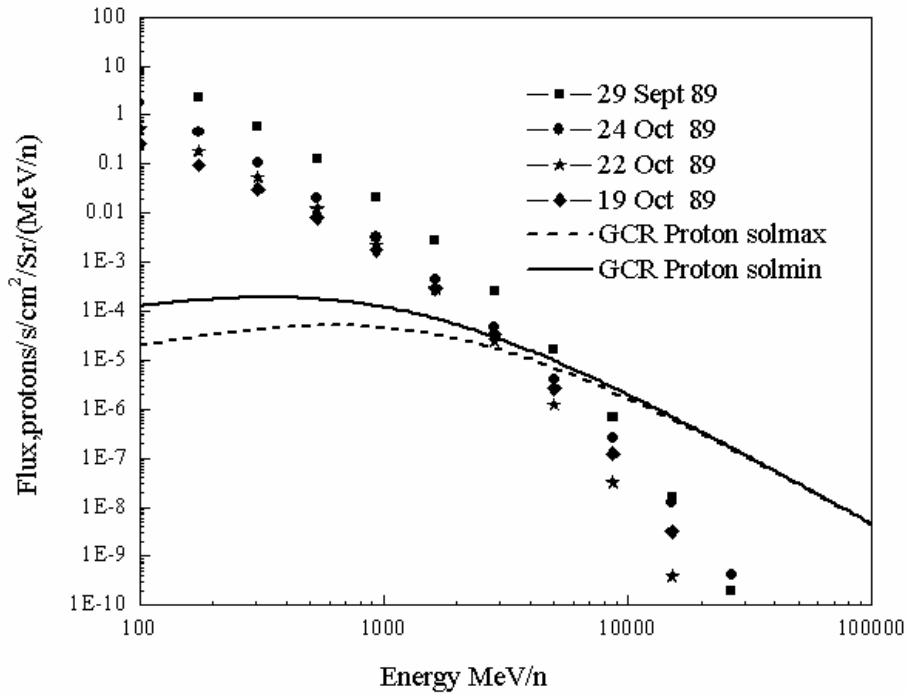

Fig. 7. Differential energy spectra at the peaks of four SEP events in 29 September,19 October, 22 October and 24october 1989 compared with cosmic ray spectrum at solar maximum and solar minimum.

## 3. Charging simulation results

We have simulated the primary GCR and SEP proton spectra in the energy range 0.1–1000 GeV and 0.1–10 GeV, respectively. Only particles with energies larger than 100 MeV/n have been considered, as below this energy the nucleonic component of cosmic rays stops inside the spacecraft without reaching the test masses. For this reason, impulsive solar events have been disregarded. The limit of 10 GeV for proton solar events has been determined by the data availability. However, the sharply decaying SEP energy spectra are very poorly populated at high energies. Therefore, their contribution to the TM charging process plays a minor role.

We have run four independent GEANT4 simulations to determine the charging of ASTROD I test mass by the four SEP event peak fluxes (29 Sept, 19 Oct, 22Oct and 24 Oct, 1989). The details of each particle event that resulted in test mass charging were recorded, including the event time and net charge deposited on the test mass. The variation of the net charge with time due to the SEP fluxes is shown in Fig. 8. The curves show the test mass net charge as a function of time obtained by simulations. The straight lines correspond to a least squares fit of these data, giving the mean net charging rates of 64323 +e/s, 2848.5 +e/s, 5201.7+e/s and 13095 +e/s, for 29 Sept, 19 Oct, 22 Oct and 24 Oct, 1989, respectively. The magnitudes of these charging rates are 300 to 7000 times larger than the charging rates due to GCR proton fluxes at solar maximum.

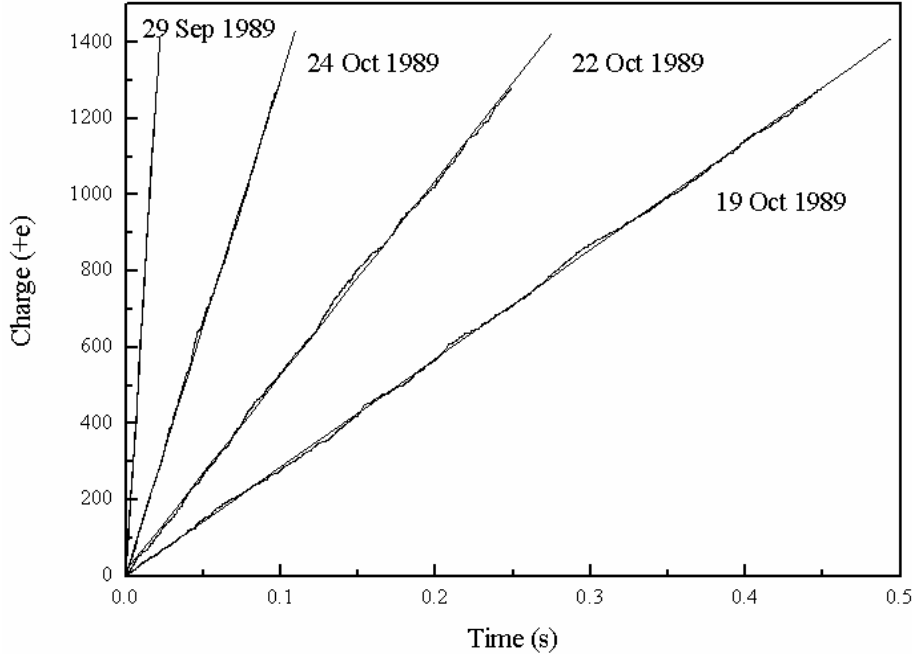

Fig. 8. Charge accrual as a function of time for four SEP events in 1989. The curves show the test mass net charge as a function of time obtained by simulations and the straight lines correspond to a least squares fit of these simulated data.

## 4. Charging disturbances

The accumulation of charge on the TM will give rise to acceleration noise in the measurement bandwidth through both Coulomb and Lorentz interactions (Shaul et al., 2005a). Further, the position dependence of Coulomb forces can modify the effective stiffness, or coupling between the test mass and the spacecraft. These disturbances are evaluated as follows.

The charge-dependent Coulomb acceleration $a_{Qk}$ in direction $\hat{k}$ is given by,

$$a_{Qk} = \frac{Q^2}{2mC_T^2}\frac{\partial C_T}{\partial k} + \frac{QV_T}{mC_T}\frac{\partial C_T}{\partial k} - \frac{Q}{mC_T}\sum_{i=1}^{N-1}V_i\frac{\partial C_{i,N}}{\partial k}, \qquad (2)$$

where $m$ is the mass of the TM; Q is the free charge accumulated on the TM; $C_{i,j}$ is the capacitance between conductors $i$ and $j$ which surround the TM; $V_i$ is the potential to which conductor $i$ is raised; $C_T$ is the capacitance coefficient of the TM. (Shaul et al., 2005a). The first two terms in Eq. (2) are dependent on the overall sensor geometric symmetry and the third term is dependent on the symmetry of the sensor voltage distribution. The corresponding acceleration noise $\delta a_{Qk}$, due to random fluctuations of the test mass position relative to the spacecraft, $\delta k$, of the potentials of the conductors that surround the test mass, $\delta V_i$, and of the test mass free charge $\delta Q$, is given by,

$$\delta a_{Qk}^2 = (\frac{\partial a_{Qk}}{\partial k})^2 \delta k^2 + \sum_{i=1}^{N-1}(\frac{\partial a_{Qk}}{\partial V_i})^2 \delta V_i^2 + (\frac{\partial a_{Qk}}{\partial Q})^2 \delta Q^2, \qquad (3)$$

We used the typical parameter values to estimate the acceleration noises: $m$ = 1.75 kg; mean voltages on opposing conductors $V_i$ = 0.5 V; the potential difference between conductors on opposing faces of the sensor compensated to 10 mV; the asymmetry in gap across opposite sides of TM = 10 μm; capacitances and capacitance gradients were calculated using parallel plate approximations: $C_T$ = 53 pF; $V_T$ = 0.5 V; position noise $\delta k$ = 1×10$^{-7}$ m Hz$^{-1/2}$; voltage noise $\delta V_i$ = 1×10$^{-4}$ V Hz$^{-1/2}$ and charge noise $\delta Q$ = 4.6×10$^{-15}$ C Hz$^{-1/2}$, which includes, as for the charging rate, the unmodelled contributions.

Lorentz effects arise from the motion of the test mass through the interplanetary magnetic field, $\vec{B}_I$, and its residual motion through the field generated within the spacecraft, $\vec{B}_S$. The Lorentz acceleration noise $\delta a_L$ is given by,

$$m^2(\delta a_L)^2 \approx (\eta Q V_I \delta B_I)^2 + (\eta Q \delta V_I B_I)^2 + (Q \delta V_S B_S)^2 + (\eta \delta Q V_I B_I)^2, \qquad (4)$$

where $V_I$ is the speed of the TM through the interplanetary field; $\delta V_I$ and $\delta V_S$ are the magnitudes of random fluctuations in the TM velocity through the interplanetary field and relative to the spacecraft, and $\delta B_I$ gives the magnitude of fluctuations in the interplanetary field (Shaul et al., 2005a). The parameters used in Eq. (4) are: $\eta$ = 0.1; $\bar{V}_I$ = 4×10$^4$ m s$^{-1}$; $\delta V_I$ = 4.78×10$^{-12}$ m s$^{-1}$ Hz$^{-1/2}$; $\delta V_S$ = 6.28×10$^{-11}$ m s$^{-1}$ Hz$^{-1/2}$; $\bar{B}_S$ = 9.6×10$^{-6}$ T; $|\delta B_S|$ = 1×10$^{-7}$ T Hz$^{-1/2}$; $\bar{B}_I$ = 1.2×10$^{-7}$ T (this is a conservative estimate of the field at 0.5AU, used to give the worst-case noise, for the ASTROD orbit); $|\delta B_I|$ = 1.2×10$^{-6}$ T Hz$^{-1/2}$.

We have calculated these two kinds of acceleration noises using Eq. (3) and Eq. (4), and summarized the results of charging disturbances due to SEPs and GCR protons in Table 1.

| Source | Flux (p/cm$^2$/s) | Charge rate (+e/s) | Coulomb Noise at 0.1mHz (m s$^{-2}$ Hz$^{-1/2}$) | Lorentz Noise at 0.1mHz (m s$^{-2}$ Hz$^{-1/2}$) |
|---|---|---|---|---|
| Primaries at solar maximum | 1.89 | 9.0 | 2.80×10$^{-15}$ | 2.80×10$^{-15}$ |
| Primaries at solar minimum | 4.29 | 26.5 | 1.40×10$^{-15}$ | 1.30×10$^{-15}$ |
| SEP peak flux of 19th Oct 1989 | 314.92 | 2848.5 | 2.11×10$^{-14}$ | 1.84×10$^{-14}$ |
| SEP peak flux of 22th Oct 1989 | 573.88 | 5201.7 | 3.37×10$^{-14}$ | 2.96×10$^{-14}$ |
| SEP peak flux of 24th Oct 1989 | 1441.5 | 13095 | 1.54×10$^{-13}$ | 1.24×10$^{-13}$ |
| SEP peak flux of 29th Sept 1989 | 7092.51 | 64323 | 4.98×10$^{-12}$ | 1.05×10$^{-12}$ |

Table 1. Estimates of integrated fluxes, charging rates and charging disturbances by four SEP events peak fluxes in 1989.

## 5. Conclusion

The charging process of the ASTROD I test mass during SEPs has been simulated using the GEANT4 toolkit. The Monte Carlo model predicted a charging rate which is much larger than that due to the GCR proton flux at solar maximum and solar minimum.

The ASTROD I acceleration noise limit target is $10^{-13}$ m s$^{-2}$ Hz$^{-1/2}$ at 0.1 mHz (the lowest frequency in the ASTROD I requirement bandwidth) (Shiomi and Ni, 2006). Comparing with the charging disturbances summarized in Table 1, the magnitudes of the Coulomb and Lorentz acceleration noise of SEP events on 19 Oct and 22 Oct are well below the acceleration noise target, however, the noise from the two larger events (29 Sept and 24 Oct) exceeds the acceleration noise budget requirement. Continuous discharge should be considered for suppressing these charging noises as considered by Shaul et al. (2005b).

In the future, we will characterize these SEP events in terms of their frequency of occurrence, duration, peak flux and total fluences as well as energy spectrum. We will also investigate whether using the discharge system to slowly accumulate opposite polarity charge for a few hours preceding SEP onset could reduce noise over the event duration (Shaul et al., 2006).

## Acknowledgments

This work is funded by the National Natural Science Foundation (Grant Nos 10475114 and 10573037) and the Foundation of Minor Planets of Purple Mountain Observatory.